# Atomic scale variation of electron tunneling into a Luttinger liquid ?
## : High resolution scanning tunneling spectroscopy study on Au/Ge(001)

Jewook Park[1], Kan Nakatsuji[3], Tae-Hwan Kim[2], Sun Kyu Song[1], Fumio Komori[4], and Han Woong Yeom[1,2*]

[1]*Center for Artificial Low Dimensional Electronic Systems, Institute for Basic Science, Pohang 790-784, Korea*
[2]*Department of Physics, Pohang University of Science and Technology, Pohang 790-784, Korea*
[3]*Department of Materials Science and Engineering, Interdisciplinary Graduate School of Science and Engineering, Tokyo Institute of Technology, Japan.*
[4]*Institute for Solid State Physics, University of Tokyo, Japan.*

Au-induced atomic wires on the Ge(001) surface were recently claimed to be an ideal 1D metal and their tunneling spectra were analyzed as the manifestation of a Tomonaga-Luttinger liquid (TLL) state. We reinvestigate this system for atomically well-ordered areas of the surface with high resolution scanning tunneling microscopy and spectroscopy (STS). The local density-of-states maps do not provide any evidence of a metallic 1D electron channel along the wires. Moreover, the atomically resolved tunneling spectra near the Fermi energy are dominated by local density-of-states features, deviating qualitatively from the power-law behavior. On the other hand, the defects strongly affect the tunneling spectra near the Fermi level. These results do not support the possibility of a TLL state for this system. An 1D metallic system with well-defined 1D bands and without defects are required for the STS study of a TLL state.

Tomonaga-Luttinger liquid (TLL) is undoubtedly one of the most important theoretical models for interacting electrons in one dimension (1D) [1-3]. Over past decades, many efforts were made to experimentally observe a TLL state. The evidence for a TLL state has been accumulated in carbon nanotubes [4, 5], strongly anisotropic bulk crystals [6-8], fractional-quantum-hall-effect edge states [9], and 1D electron gases of quantum wires [10].

Along a largely different direction, the possibility of a TLL state was also discussed in metallic atomic wires self-organized on semiconductor surfaces, in particular, for Au-induced atomic wire arrays on vicinal silicon surfaces [11]. However, no clear indication of a TLL state has been identified for these systems so far [12-15]. As the most recent system in this line of researches, Au-induced atomic wires on the Ge(001) surface (hereafter, the Au-Ge wires) were suggested as an ideal 1D metallic system [16], with a clear signature of a TLL state in their scanning tunneling spectroscopy (STS) spectra [17]. However, not only the chemical composition and the atomic structure [18-24], but also their band structure is uncertain at present [25, 26]. Most notably, a recent angle-resolved photoemission spectroscopy (ARPES) study, the first such study for a single domain surface, showed an anisotropic but 2D metallic band, which disperses more strongly in the direction perpendicular to the wire [27, 28].

This situation apparently and urgently requests the confirmation of the existence of a 1D metallic state itself in the Au-Ge wire and the TLL behavior of its tunneling spectra. We also note that while the importance of STS has been mentioned for a few 1D metallic systems [7, 29], no detailed atomic scale investigation of STS spectra of a TLL system is available. For such an atomic scale study, a well ordered surface 1D metallic system would definitely be beneficial. Therefore, the Au-Ge wire could be an important model system to unveil a largely

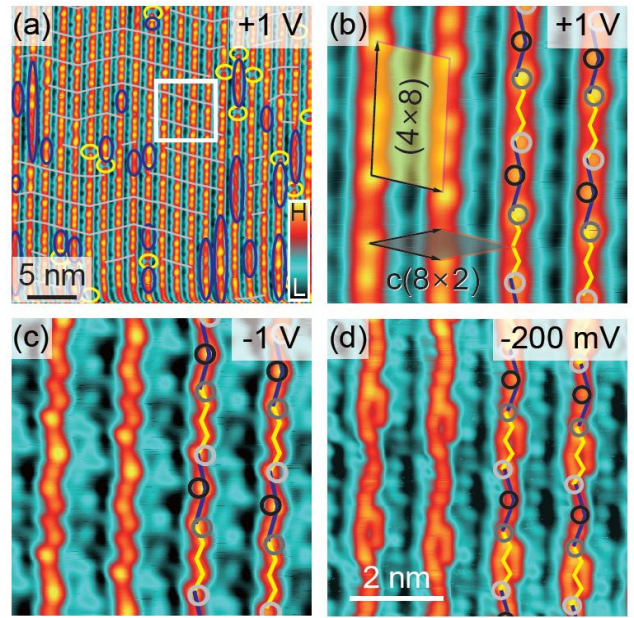

FIG 1 (a) STM topography image of the Au-Ge wires ($33 \times 33$ nm$^2$, tunneling current $I_t = 50$ pA). Gray lines connect neighboring ($4\times8$) units to show the irregularity (blue and yellow ovals). (b)-(d) Enlarged bias dependent topographies ($I_t = 100$ pA) on the perfectly ordered area [$6.7 \times 6.7$ nm$^2$, the squared area in (a)]. The circles (gray, black, and dark gray) together with 'V' (dark blue) and 'W' (yellow) lines are overlaid in order to show the schematics of the protrusions; the three circles from the empty (+1 V) state image and V (W) shapes from filled (−1 V) state images. The ($4\times8$) and $c(8\times2)$ unit cells are shown in (b).

unexplored area, the atomic scale tunneling properties of a TLL.

In this paper, we scrutinized the Au-Ge wires using high resolution scanning tunneling microscopy (STM) and STS. In particular, we focused on atomically perfectly ordered parts of the surface in order to avoid notorious effects of defects on a TLL state. Our well resolved STS maps show no clear indication of a 1D electron channel around the Fermi energy ($E_F$), which is consistent with the recent ARPES study [28]. Moreover, we found that the detailed STS spectra near the $E_F$ are not governed by the power-law behavior of a TLL but by strong density-of-states features and defect states. A model surface system to host ideally 1D metallic electrons remains to be found.

We prepared the Au-Ge wires based on the Au coverage and growth conditions well established in the previous studies [30-32]. We prepared a clean and well-ordered Ge(001) substrate (Sb-doped, 0.2-0.4 Ωcm) with several cycles of 1 keV Ar$^+$ sputtering, annealing (950 K), and flash heating (1050 K) in ultra-high vacuum. Subsequently, we deposited 0.75 monolayer of Au on the substrate held at 770 K using a thermal evaporator. We used a commercial cryogenic STM (Unisoku, Japan) to perform STM and STS measurements at 78 K and 5.5 K. STS data were taken by the standard lock-in technique with a bias voltage modulation of 500 Hz and 4-30 mV.

We first investigated the STM topography of the Au/Ge(001) surface for a wide terrace area (width > 400 nm) at 78 K. The bias-dependent topography images in Fig. 1 display the wire array with a regular width and a constant interwire spacing of 1.6 nm ($4a_0$ with $a_0$ = 4 Å, the unit cell size of the substrate surface either in the [110] or [$\bar{1}$10] direction) [33, 34]. As shown in Fig. 1(b), within a well ordered region, the surface has a 4×8 unit cell with the characteristically modulated protrusions along a wire, which were previously labeled as 'V (chevron)' and 'W (zigzag)' protrusions [34, 35]. This is in clear contrast to the $c(8\times2)$ unit cell suggested initially [17], which is now thought to correspond to a higher temperature phase [34]. This group reassigned the low temperature phase as the $p(4\times1)$ superstructure of $c(8\times2)$ but is the same as the present 4×8 structure [24]. The topography images at various biases in Fig. 1 are fully consistent with the recent works [24, 33-35] while the underlying atomic structure is largely uncertain [30]. In particular, the complicated protrusions revealed at low biases [Fig. 1(d)] are not compatible with the most recent structural model [30]. Note also that the 4×8 long range order is perturbed by the frequent occurrence of intrawire defects (yellow and blue ovals) and interwire misfits (guided by gray lines) [Fig. 1(a)]. The well ordered 4×8 patches extend only to roughly 10×10 nm$^2$ at maximum and we mainly focus on such ordered patches [see the square in Fig. 1(a)] to rule out any unwanted disorder effects. This degree of disorder is consistent over most of the STM images reported by several different groups [20, 31, 33, 35].

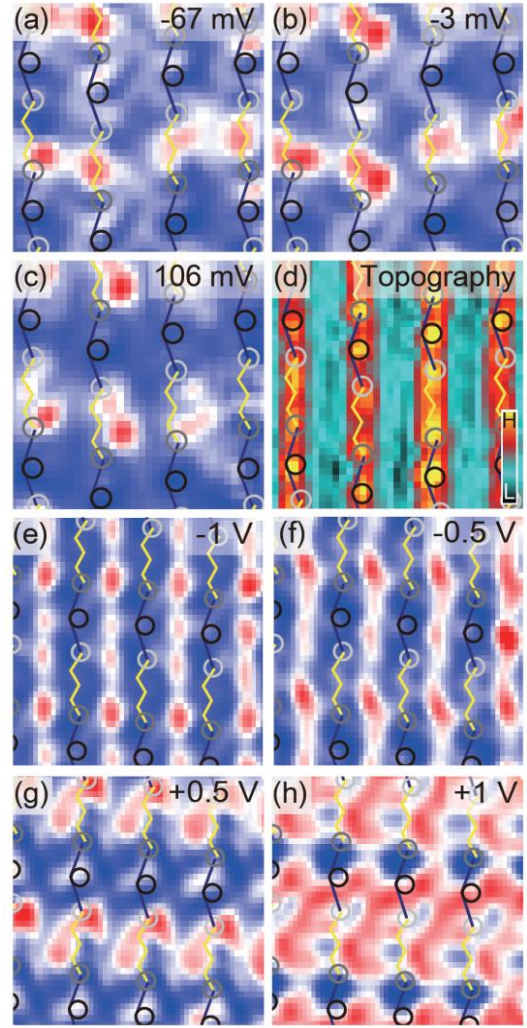

FIG 2 (a)-(c) Bias dependent differential conductance ($dI/dV$) maps at low biases with (d) simultaneously obtained topography image (6 × 6 nm$^2$, $V_b$ = 0.5 V, $I_t$ = 200 pA, 78 K). The three circles and the V(W) shapes guide the spatial variation of LDOS features within the spectroscopic maps. (d)-(j) Similar maps at higher biases (5.9 × 5.9 nm$^2$, $V_b$ = 1 V, $I_t$ = 100 pA, 78 K) from the other ordered region shown in Fig. 1. Color scale of $dI/dV$ map fits to each image.

Local density of states (LDOS) (STS $dI/dV$) maps are essential in addressing the existence of a metallic electron channel. A recent LDOS measurement showed an electronic channel along the *trenches* between the Au-Ge wires at −100 meV [37]. This is in sharp contrast with the original claim of a high conductance 1D channel *on* the Au-Ge wires, which was based not on the STS data but only on the topography image [21]. A more recent STS study claimed 1D channel *on* the Au-Ge wires at ±20 meV, suggesting again the 1D metallic nature but with a strong inhomogeneity due possibly to defects [24].

Figure 2 shows an example of such low bias $dI/dV$ maps obtained on a perfectly ordered 4×8 patch. We acquired a spatially resolved LDOS map at a given bias by slicing a full grid STS data. Above −200 meV, we can

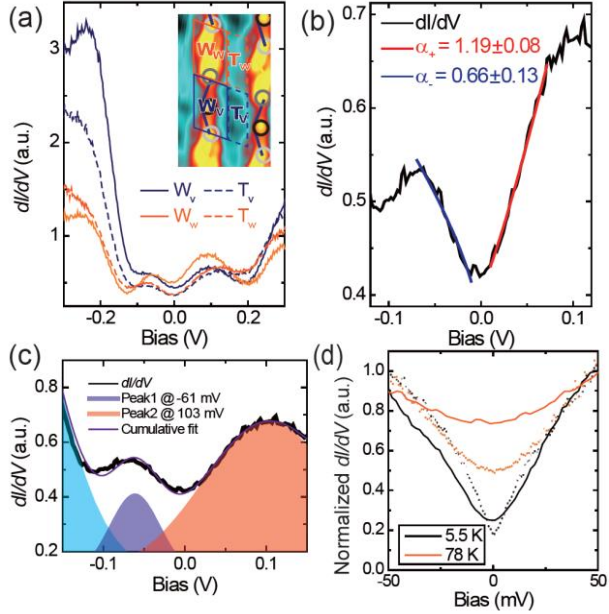

FIG 3 (a) Position dependent *dI/dV* conductance curve from a 4×8 structure in Fig. 2(e) ($V_b$ = 300 mV, $I_t$ = 200 pA, 78 K). The solid and dashed lines correspond to the *dI/dV* data from wire ($W_w$ and $W_v$) and trench regions ($T_w$ and $T_v$), respectively, shown in the inset. Each spectrum is averaged over the data points within each box of the inset. (b) Area-averaged *dI/dV* conductance curve of 4 × 8 structure. The power-law fitting results near $E_F$ is overlaid for empty and filled states, separately, in red and blue solid lines, respectively. $\alpha_+$ ($\alpha_-$) is power exponent for empty (filled) states. (c) The *dI/dV* curve near Fermi level with peak fitting. Asymmetric peaks at −61 mV (blue) and 103 mV (red) govern electronic properties near Femi level. Cyan colored peak originated from LDOS peak at −262 mV (d) LDOS feature near Fermi level. Black and orange solid lines are normalized *dI/dV* conductance at 5.5 K and 78 K, respectively. Dimmed dotted curves are corresponding 4.7 K and 78 K data from Fig.3a of ref [17].

confirm the existence of a 1D electron channel running along the trenches between the wires as observed recently [36]. This feature, but, becomes weaker at a lower bias and almost vanishes at $E_F$. In addition, there exists a strong LDOS feature around the so called 'W' shapes on the wires and this feature dominates at lower biases by showing regular periodicity along parallel as well as perpendicular direction to the Au-Ge wire. However, one cannot find any clear evidence of the existence of any 1D metallic channel along the wires against the impression provided by the topography images at low bias. This discrepancy between the topography and LDOS maps was suggested to originate from the strong height corrugation across the wires [20, 34]. Judged from the above data, we most reasonably conclude that a 1D metallic channel is not formed on this surface. This is fully consistent with the recent ARPES result [28], which showed a 2D metallic band, dispersing more strongly *across* the wires with even a gap near the $E_F$ *along* the wire.

Given the absence of a well-defined 1D metallic electron channel along the wire, it is hard to understand the TLL power-law behavior of STS spectra [17]. Figure 3(a) shows the detailed *dI/dV* curve averaged on four inequivalent parts (solid and dashed boxes in the inset) of a 4×8 unit cell away from any defect. This avoids the inclusion of possible disorder effects and the erroneous sampling of point-by-point STS spectra. The spectra consist of four characteristic peaks at −262, −61, +103, and +303 meV, which has strong and systematic lateral dependences as shown in this figure and Fig. 2.

While the previous studies could not resolve these LDOS peaks (at −61 and +103 meV) near the $E_F$, next nearest peaks (at −262 and 303 meV) are largely consistent [25, 33]. As shown below, these LDOS features are suppressed for bad tip conditions as well as defective regions, and are largely wiped out for lateral averaging including various defects as performed in the previous study [17]. The origin of the LDOS peaks, at least for the filled states, can be traced in the band structure probed by ARPES studies. In particular, the peak at −61 meV could correspond to the bottom of the metallic band found around −100 meV in ARPES [27]. The empty state peak at +103 meV can reasonably be related to the top of this band, whose band width is then estimated as 170-200 meV.

Figure 3(b) enlarges the 'raw' *dI/dV* curve near the $E_F$. This exhibits a large asymmetry between the filled and empty states and the power law fit gives largely different power exponent values for empty ($\alpha_+$ = 1.19) and filled ($\alpha_-$ = 0.66) state. The filled state power exponent value is roughly consistent with the previous report [17]. The asymmetry of the STS spectra across $E_F$ can also be noticed in another report of the same group [33] and in the other group's report [25]. This power exponent value for the spatially well resolved STS spectra also exhibit systematic dependence according to the surface superstructure as can be easily noticed in Figs. 3(a) and 4(b). These spectral characteristics are qualitatively different from the TLL behavior. As shown in Fig. 3(c), this is due to fact that the STS spectra near $E_F$ is dominated by the finite width LDOS peaks at −61 and +103 meV, which has strong lateral dependence. That is, the dip structure at $E_F$ [Fig. 3(d)] is apparently not due to the power-law suppression of the LDOS but is defined by the two LDOS peaks and their widths.

The remaining question is what caused the apparent power-law-like decay in the previous STS data. The most probably cause is the disorder [37], which is indeed substantial in the present system. As shown in Fig. 4, this surface has various atomic scale defects and these defects have strong spectroscopic footprint near $E_F$ in particular at the filled state. Most of these disorders were unfortunately ignored in the previous spatially averaged STS study [17] while we count a similar number of defects in the STM images of this study. Figure 4 shows apparently that the averaging including defects suppresses the peak at −61 meV significantly since each defect has its own LDOS

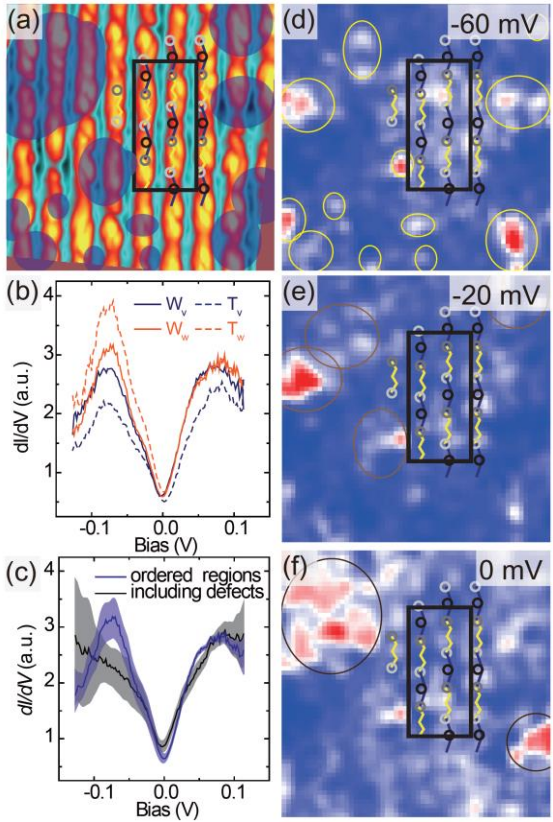

FIG 4 (a) STM topography image and $dI/dV$ conductance curves at 5.5 K (15.2 × 15.2 nm$^2$, $V_b$ = 0.1 V, $I_t$ = 200 pA). Shaded ovals are where defect-related states are observed in corresponding $dI/dV$ maps (d-f). (b) $dI/dV$ conductance curves from defect-free region. Curves are acquired by averaging a few W$_V$, W$_W$, T$_V$, and T$_W$ regions in black box of (a), separately. Each curves are selected as parallelograms likely to Fig. 3(a). (c) Blue and black area-averaged $dI/dV$ curves are acquired from black box in (a) (blue) and overall field of view of (a) (black) including defect-related regions, respectively. Shade of each curve means error bar. (d)-(f) Bias dependent $dI/dV$ conductance map corresponds to (a). Local defect-induced features are marked as ovals.

peak energy at a different energy as shown in Fig. 4. This situation can be exaggerated depending on the tip condition, where the transmission near $E_F$ can be easily reduced further. We believe that the previous study averaged point spectra for a rather large area including unavoidably various defects, which wipes out the LDOS peak signature and makes the structure around $E_F$ more like an anomalous dip as in Fig. 4(c) [17]. This emphasizes the importance of disorder and the need for a proper STS study for a defect-free system with a reproducible and reliable tip condition. The defects/disorder effect can also explain the power-law like ARPES intensity near $E_F$ and such an effect is strongly suggested by the extraordinarily board spectral width of the surface state band in both momentum and energy distributions [25-28]. We may suggest that the scaling behavior of the STS spectra observed previously [17] can simply be due to the systematic temperature broadening of the two LDOS peaks defining the dip at the Fermi energy.

In conclusion, we showed that the atomic wire structure on the Au/Ge(001) surface has a largely 2D electronic state near $E_F$ and its STS spectra can be explained by the LDOS features. This denies the previous claim of an ideal 1D metal and the TLL behavior for this system. A 1D metallic state of atomic wires on semiconductor surfaces remains to be found, in which the atomic scale tunneling properties of a TLL could be disclosed.

This work was supported by Institute for Basic Science (IBS-R014-D1). KN and FK were supported by KAKENHI (21244048 and 22710095) from JSPS, Japan, and by the NSG Foundation.

[*]yeom@postech.ac.kr